\documentstyle[11pt,fleqn]{article}

\newcommand{\be}{\begin{eqnarray}}
\newcommand{\ee}{\end{eqnarray}}

\def\ap[#1,#2]{a_#1 + a_#2}
\def\a[#1,#2]{a_#1 - a_#2}

\begin{document}
%%%%%%%%%%%%%%%%%%%%%%%%%%%%%%%%%%%%%%%%%%%%%%%%%%%%%%%%%
\title{Free energies and critical exponents of the\\ 
$A_1^{(1)}, B_n^{(1)}, C_n^{(1)}$ and $D_n^{(1)}$ face models} 
\author{M. T. Batchelor$^{\rm a}$, V. Fridkin$^{\rm a}$, 
A. Kuniba$^{\rm b}$, K. Sakai$^{\rm b}$ and Y.-K. Zhou$^{\rm a}$\\\\
$^{\rm a}$ Department of Mathematics, School of Mathematical Sciences,\\ 
Australian National University, Canberra ACT 0200, Australia\\
$^{\rm b}$ Institute of Physics, University of Tokyo, Komaba, Meguro-ku,\\ 
Tokyo 153, Japan}

\maketitle
%%%%%%%%%%%%%%%%%%%%%%%%%%%%%%%%%%%%%%%%%%%%%%%%%%%%%%%%%
\pagenumbering{arabic}
\begin{abstract}
%%%%%%%%%%%%%%%%%%%%abstract
We obtain the free energies and critical exponents of models
associated with elliptic solutions of the star-triangle relation and
reflection equation.  The models considered are related to the affine
Lie algebras $A_1^{(1)}$, $B_n^{(1)},C_n^{(1)}$ and $D_n^{(1)}$. The
bulk and surface specific heat exponents are seen to satisfy the
scaling relation $2\alpha_s = \alpha_b + 2$. It follows from 
scaling relations that in regime III the correlation length exponent 
$\nu$ is given by $\nu=(l+g)/2g$, where $l$ is the level and $g$ is the 
dual Coxeter number. In regime II we find $\nu=(l+g)/2l$.
\end{abstract}


\vskip 5 mm
%\newpage
%%%%%%%%%%%%%%%%%%%%%%%%%%%%%%%%%%%%%%%%%%%%%%%%%%%%%%%%%
%  The start of the main text
%%%%%%%%%%%%%%%%%%%%%%%%%%%%%%%%%%%%%%%%%%%%%%%%%%%%%%%%%

The formulation of the boundary version of the Yang-Baxter equation
has provided a systematic framework for the investigation of
integrable models with a boundary [1-3]. 
Recent attention has turned to exploiting boundary integrability
to derive off-critical surface phenomena, such as the surface 
magnetization \cite{jkkkm95,jkkmw95},
the surface free energy 
%and related critical exponents \cite{z96,bz96,zb96,zb96b,bfz96,opb95}.
and related critical exponents [6-11].
For face models,  
integrability in the bulk is assured by solutions of the star-triangle 
relation (STR) \cite{baxter}
%%%%%%%%%%%%%%%%%%%%%%%%%%%%%%%%%%%%%%%%%%%%%%%%%%%%%%%%%
%               Begin Definitions 
%%%%%%%%%%%%%%%%%%%%%%%%%%%%%%%%%%%%%%%%%%%%%%%%%%%%%%%%%
%
% Term representations
%
\def\W[#1,#2,#3,#4,#5]{ W \left( \begin{array}{@{}cc|} 
                        #1 & #4 \\ #2 & #3 \end{array} \; #5 \right)}
\def\Wp[#1,#2,#3,#4]{ \W [a,a+#1,#2,a+#4, u]} 
\def\K[#1,#2,#3,#4] { K \left( \begin{array}{@{}cc|}
                         & #3 \\ \raisebox{1.2ex}{$#1$}  & #2 \end{array} \; #4 \right)}
%
%%%%%%%%%%%%%%%%%%%%%%%%%%%%%%%%%%%%%%%%%%%%%%%%%%%%%%%%%
%               End Definitions 
%%%%%%%%%%%%%%%%%%%%%%%%%%%%%%%%%%%%%%%%%%%%%%%%%%%%%%%%%
\be \begin{array}{l}
\displaystyle{\sum_g} \W[f,a,g,e,u] \W[a,b,c,g,v] \W[e,g,c,d,u-v] = \\[3.0ex] 
\hphantom{\W[f,a,g,e,u]}
\displaystyle{\sum_g} \W[f,a,b,g,u-v] \W[g,b,c,d,u] \W[f,g,d,e,v].
   \end{array}
\label{str}
\ee
Integrability at a boundary requires the additional relation,
\be \begin{array}{l} \displaystyle{\sum_{fg}} \,\,
\W[c,g,a,b,u-v] \K[g,f,a,u] \W[c,d,f,g,u+v] \K[d,e,f,v] = \\[3.0ex]
\hphantom{WW} \displaystyle{\sum_{fg}} \,\,
\K[b,f,a,v] \W[c,g,f,b,u+v] \K[g,e,f,u] \W[c,d,e,g,u-v],
   \end{array}
\label{re}
\ee
which is the face formulation of the reflection 
equation (RE) \cite{bpo96,fhs95,z96,ku96,ak96}. 

Solutions to the RE, defining boundary weights, have recently 
been found for the $A_n^{(1)}$, $A_n^{(2)}$ and 
$X_n^{(1)}=B_n^{(1)},C_n^{(1)},D_n^{(1)}$ models \cite{bfkz96}, 
for which the solutions of the STR
have been known for some time \cite{jmo88,k91}.
Our interest here lies in the critical behaviour of the bulk and surface free
energies of the elliptic face models associated with 
the algebras $A_1^{(1)}$ and $X_n^{(1)}$. 
These models include some well known models as
special cases. For example, 
the Andrews-Baxter-Forrester (ABF) model \cite{abf84} 
is related to $A_1^{(1)}=C_1^{(1)}$.

%%%%%%%%%%%%%%%%%%%%%%%%%%%%%%%%%%%%%%%%%%%%%%%%%%%%%%%%%%%%
%               Begin Definitions
%%%%%%%%%%%%%%%%%%%%%%%%%%%%%%%%%%%%%%%%%%%%%%%%%%%%%%%%%%%%
\def\twosquare
{\begin{picture}(80,40)(0,0)
\put(0,20){\line(1,1){20}}
\put(0,20){\line(1,-1){20}}
\put(20,0){\line(1,1){40}}
\put(20,40){\line(1,-1){40}}
\put(80,20){\line(-1,1){20}}
\put(80,20){\line(-1,-1){20}}
\end{picture}}
\def\inversionA
{\begin{picture}(80,40)(0,0)
\put(0,20){\line(1,1){20}}
\put(0,20){\line(1,-1){20}}
\put(20,0){\line(1,1){40}}      % Two squares
\put(20,40){\line(1,-1){40}}
\put(80,20){\line(-1,1){20}}
\put(80,20){\line(-1,-1){20}}
\put(15,35){\line(1,0){10}}
\put(55,35){\line(1,0){10}}     % Angles
\put(40,20){\circle*{4}}
\put(20,20){\makebox(0,0){$u$}}         % Spectral params
\put(60,20){\makebox(0,0){$-u$}}
\multiput(30,5)(5,0){5}{\circle*{1}}    % identification
\multiput(30,35)(5,0){5}{\circle*{1}}
\end{picture}}
\def\inversionB
{\begin{picture}(80,40)(0,0)
\put(0,20){\line(1,1){20}}
\put(0,20){\line(1,-1){20}}
\put(20,0){\line(1,1){40}}      % Two squares
\put(20,40){\line(1,-1){40}}
\put(80,20){\line(-1,1){20}}
\put(80,20){\line(-1,-1){20}}
\put(45,15){\line(0,1){10}}
\put(35,15){\line(0,1){10}}     % Angles
\put(40,20){\circle*{4}}
\put(20,20){\makebox(0,0){$\lambda-u$}}         % Spectral params
\put(60,20){\makebox(0,0){$\lambda+u$}}
\multiput(30,5)(5,0){5}{\circle*{1}}    % identification
\multiput(30,35)(5,0){5}{\circle*{1}}
\end{picture}}
%%%%%%%%%%%%%%%%%%%%%%%%%%%%%%%%%%%%%%%%%%%%%%%%%%%%%%%%%
%            End Definitions
%%%%%%%%%%%%%%%%%%%%%%%%%%%%%%%%%%%%%%%%%%%%%%%%%%%%%%%%%
Two inversion relations,
\be \displaystyle{\sum_g}\;
 \W[a,b,c,g,u] \W[a,g,c,d,-u] &=& \delta_{bd} \varrho(u), \\
 \displaystyle{\sum_g}
\left( {\displaystyle\frac{G_g G_b}{G_a G_c}} \right)
        \W[g,a,b,c,\lambda-u] \W[g,c,d,a,\lambda+u] &=& \delta_{bd}
\varrho(u),
\ee
are satisfied by the bulk weights of the models under consideration.
On the other hand, the diagonal solutions of the RE found in \cite{bfkz96}
fulfill the boundary crossing relation 
\be
\sum_g \left( {\displaystyle\frac{G_g}{G_b}} \right)^{1/2}
\W[a,b,c,g,2u+\lambda] \K[g,c,a,u+\lambda] = \varrho_s(u) \K[b,c,a,-u] .
\label{bcr}
\ee
The $G_a$ are crossing factors \cite{jmo88}.
The crossing parameter is given by $\lambda = -tg/2$ where the 
parameters $t,g$ are given in Table 1.
The key ingredients are the inversion function $\varrho(u)$ and the 
boundary crossing function $\varrho_s(u)$.

The inverson relation method \cite{baxter} has recently been
applied to a number of
%models to obtain the surface free energy \cite{bz96,zb96,zb96b,bfz96,opb95}.
models to obtain the off-critical surface free energy [7-11].
The unitarity relation
\begin{equation}
 T(u)T(u+\lambda)=
 \frac{\varrho_s(u)\varrho_s(-u)}{\varrho(2u)}\
 \varrho^{2N}(u),
\label{eqn:a1}
\end{equation}
for the transfer matrix eigenvalues $T$($u$)
follow from the crossing unitarity relation and
disregarding finite-size corrections. Define
$T_b(u)=\kappa_b^{2N}$ and
$T_s(u)=\kappa_s$, then the bulk
and the surface free energies per site can be defined by
$f_b(u)=-\log\kappa_b(u)$ and $f_s(u)=-\log\kappa_s(u)$,
respectively.

The restricted-solid-on-solid (RSOS) models follow in a natural way
from the unrestricted models that we have discussed so far.
One introduces a positive integer $l$ and sets $L$
as specified in Table 1.
Local state $a$ in the Boltzmann weights is taken as a level
$l$ dominant integral weight of $A^{(1)}_1$ and $X^{(1)}_n$.
In Table 1 we have also listed the levels under consideration.
We do not treat $l=1$ for the $B^{(1)}_n, D^{(1)}_n$ RSOS models
as they are then completely frozen.
%
%%%%%%%%%%%%%%%%%%%%%%%%%%%%%%%%%%%%%%%%%%%%%%%%%%%%%%%%%

We begin with the $A^{(1)}_1$ model, for which
\begin{equation}
  \varrho(u)=
  \frac{[1+u][1-u]}{[1]^2}\
\end{equation}
and
\begin{equation}
  \varrho_s(u)=
  \frac{[2-2u]}{[1]}.\
\end{equation}
Here we define
\be
[u] = [u,p] = \vartheta_1(\pi u /L,p),
\ee
where
\be
\vartheta_1(u,p) = 2 p^{1/8} \sin u {\displaystyle\prod_{n=1}^\infty}
(1 - 2p^n \cos 2u + p^{2n}) (1-p^n)
\ee
is a standard elliptic theta-function of nome $p = e^{2\pi i\tau}$.
We first consider the model in regime III ($-1<u<0$ and $0<p<1$).

After taking a convenient normalisation in (\ref{eqn:a1}) we have
\begin{equation}
  \kappa_b(u)\kappa_b(-1+u)=
  \frac{[1+u][1-u]}{[1]^2}\
\label{eqn:inversion}
\end{equation}
for the bulk and
\begin{equation}
  \kappa_s(u)\kappa_s(-1+u)=
  \frac{[2+2u][2-2u]}{[2]^2}\
\end{equation}
for the surface.
To proceed, we introduce the new variables
\begin{equation}
  x=e^{-4\pi^2 \lambda/\varepsilon}, \,
  w=e^{-4\pi^2 u/\varepsilon}, \,
  q=e^{-4\pi^2 L/\varepsilon},
\label{eqn:a6}
\end{equation}
where $\lambda = -1$ for the $A_1^{(1)}$ model. The
conjugate modulus transformation of the theta-function,
\begin{equation}
\vartheta_1(\pi u/L,p)\sim E(w,q),
\label{eqn:a7}
\end{equation}
is also required, where $p=e^{-\varepsilon/L}$ and
\begin{equation}
  E(z,y)=
 \prod_{n=1}^\infty(1-y^{n-1}z)(1-y^nz^{-1})(1-y^n).
\label{eqn:a9}
\end{equation}
We suppose that $\kappa_b(w)$ is analytic and
nonzero in the annulus $1 \le w \le x$ and
Laurent expand $\log \kappa_b(w)$ in powers of $w$.
Then matching coefficients in (\ref{eqn:inversion}) we obtain
\begin{equation}
f_b(u)=\sum_{n=- \infty}^\infty\frac
      {\sinh(\frac{2\pi^2nu}\varepsilon)
       \sinh\left(\frac{2\pi^2n\left(1+u\right)}
                  \varepsilon\right)
       \cosh\left(\frac{2\pi^2n\left(L-2\right)}
                  \varepsilon \right)}
      {n\sinh(\frac{2\pi^2nL}\varepsilon)
        \cosh(\frac{2\pi^2n}\varepsilon)}
\end{equation}
for the bulk free energy. 
In a similar manner, we obtain the surface free energy
\begin{equation}
  f_s(u)=\sum_{n=-\infty}^\infty\frac
          {\sinh(\frac{4\pi^2nu}\varepsilon)
           \sinh\left(\frac
          {4\pi^2n\left(1+u\right)}\varepsilon\right)
          \cosh\left(\frac{2\pi^2n\left(L-4\right)}
          \varepsilon \right)}
         {n\sinh(\frac{2\pi^2nL}\varepsilon)
          \cosh(\frac{4\pi^2n}\varepsilon)} .
\end{equation}

Now consider the $A_1^{(1)}$ model in 
regime II ($0<u<-1+L/2$ and $0<p<1$). In this case we need to
modify for the appropriate analyticity strip, with 
\begin{eqnarray}
\kappa_b(u)\kappa_b(-u)&=&\frac{[1+u][1-u]}{[1]^2} \\
\kappa_b(u)\kappa_b(L-2-u)&=&\frac{[2+u][u]}{[1]^2}
\end{eqnarray}
for the bulk and
\begin{equation}
  \kappa_s(u)\kappa_s(-1+L/2+u)=
  \frac{[2+2u][2-2u]}{[2]^2}\
\end{equation}
for the surface.
We assume that $\kappa_b(u)$ and $\kappa_s(u)$ are analytic and
nonzero in this regime, and in a similar manner obtain 
\begin{eqnarray}
f_b(u) &=& -\sum_{n=- \infty}^\infty\frac
              {\sinh(\frac{2\pi^2nu}\varepsilon)
              \sinh\left(\frac{2\pi^2n\left(L-3\right)}
                          \varepsilon\right)
               \sinh\left(\frac{2\pi^2n\left(L-1-u\right)}
                         \varepsilon \right)}
              {n\sinh(\frac{2\pi^2nL}\varepsilon)
                \sinh\left(\frac{2\pi^2n\left(L-2\right)}
                         \varepsilon \right)} \nonumber \\
       & & {} + \sum_{n=- \infty}^\infty\frac
              {\sinh(\frac{2\pi^2nu}\varepsilon)
               \sinh(\frac{2\pi^2n}\varepsilon)
               \sinh\left(\frac{2\pi^2n\left(1+u\right)}
                         \varepsilon \right)}
              {n\sinh(\frac{2\pi^2nL}\varepsilon)
               \sinh\left(\frac{2\pi^2n\left(L-2\right)}
                         \varepsilon \right)},
\end{eqnarray}
\begin{equation}
  f_s(u)=-\sum_{n=-\infty}^\infty\frac
          {\sinh(\frac{4\pi^2nu}\varepsilon)
           \sinh\left(\frac
          {2\pi^2n\left(L-2-2u\right)}\varepsilon\right)
          \cosh\left(\frac{2\pi^2n\left(L-4\right)}
          \varepsilon \right)}
         {n\sinh(\frac{2\pi^2nL}\varepsilon)
          \cosh\left(\frac{2\pi^2n\left(L-2\right)}
          \varepsilon\right)}
\end{equation}
for the bulk and surface free energy.

%%%%%%%%%%%%%%%%%%%%%%%%%%%%%%%%%%%%%%%%%%%%%%%%%%%%%%%%%%%%%
  
Now consider the $B_n^{(1)}$ and $D_n^{(1)}$ models in 
regime III ($\lambda<u<0$ and $0<p<1$) with $\lambda$ 
as given in Table 1. For these models the inversion and 
bounday crossing functions are given by 
\begin{equation}
  \varrho(u)=
  \frac{[\lambda+u][\lambda-u][1+u][1-u]}
  {[\lambda]^2[1]^2},
\end{equation}
\begin{equation}
  \varrho_s(u)=
  \frac{[2\lambda+2u][1-\lambda-2u]}{[\lambda][1]}.
\end{equation}
After appropriate normalization, we have
\begin{equation}
  \kappa_b(u)\kappa_b(\lambda+u)=
  \frac{[-\lambda+u][-\lambda-u][1+u][1-u]}
  {[-\lambda]^2[1]^2}
\end{equation}
for the bulk and
\begin{equation}
  \kappa_s(u)\kappa_s(\lambda+u)=
  \frac{[-2\lambda+2u][-2\lambda-2u]
  [1-\lambda+2u][1-\lambda-2u]}
  {[-2\lambda]^2[1-\lambda]^2}
\end{equation}
for the surface. Under the appropriate analyticity assumptions we 
obtain the bulk and surface free energies
\begin{equation}
f_b(u)=-2\sum_{n=- \infty}^\infty\frac
        {\sinh(\frac{2\pi^2nu}\varepsilon)
        \sinh\left(\frac{2\pi^2n\left(\lambda-u\right)}
                  \varepsilon\right)
       \cosh\left(\frac{2\pi^2n\left(L+\lambda-1\right)}
                  \varepsilon \right)
      \cosh\left(\frac{2\pi^2n\left(\lambda+1\right)}
                  \varepsilon \right)}
      {n\sinh(\frac{2\pi^2nL}\varepsilon)
        \cosh(\frac{2\pi^2n\lambda}\varepsilon)},
\end{equation}
\begin{equation}
  f_s(u)=-2\sum_{n=-\infty}^\infty\frac
          {\sinh(\frac{4\pi^2nu}\varepsilon)
          \sinh\left(\frac{4\pi^2n\left(\lambda-u\right)}
          \varepsilon\right)
          \cosh\left(\frac{2\pi^2n\left(L+3\lambda-1\right)}
          \varepsilon \right)
          \cosh\left(\frac{2\pi^2n\left(\lambda+1\right)}
          \varepsilon \right)}
         {n\sinh(\frac{2\pi^2nL}\varepsilon)
          \cosh(\frac{4\pi^2n\lambda}\varepsilon)}.
\end{equation}

%%%%%%%%%%%%%%%%%%%%%%%%%%%%%%%%%%%%%%%%%%%%%%%%%%%%%%%%%%%%%

Now consider the $B_n^{(1)}$, $C_n^{(1)}$ and $D_n^{(1)}$ models in
regime II ($0<u<\lambda+L/2$ and $0<p<1$). Similar to the $A_1^{(1)}$ 
model the inversion relations are modified to
\begin{eqnarray}
  \kappa_b(u)\kappa_b(-u)&=&
  \frac{[-\lambda+u][-\lambda-u][1+u][1-u]}
        {[-\lambda]^2[1]^2},\\
  \kappa_b(u)\kappa_b(L+2\lambda-u)&=&
  \frac{[u][-2\lambda+u][1-\lambda+u][-1-\lambda+u]}
       {[-\lambda]^2[1]^2},\\
  \kappa_s(u)\kappa_s(\lambda+L/2+u)&=&
  \frac{[-2\lambda+2u][-2\lambda-2u][1-\lambda+2u][1-\lambda-2u]}
       {[-2\lambda]^2[1-\lambda]^2}.
\end{eqnarray}
{}From these relations we obtain
\begin{eqnarray}
f_b(u) &=& -2\sum_{n=- \infty}^\infty\frac
              {\sinh(\frac{2\pi^2nu}\varepsilon)
               \cosh\left(\frac{2\pi^2n\left(\lambda+1\right)}
                          \varepsilon\right)
              \sinh\left(\frac{2\pi^2n\left(L+\lambda-u\right)}
                          \varepsilon\right)
               \sinh\left(\frac{2\pi^2n\left(L+2\lambda-1\right)}
                         \varepsilon \right)}
              {n\sinh(\frac{2\pi^2nL}\varepsilon)
                \sinh\left(\frac{2\pi^2n\left(L+2\lambda\right)}
                         \varepsilon \right)} \nonumber \\
       & & {} -2\sum_{n=- \infty}^\infty\frac
              {\sinh(\frac{2\pi^2nu}\varepsilon)
               \cosh\left(\frac{2\pi^2n\left(\lambda+1\right)}
                          \varepsilon\right)
               \sinh\left(\frac{2\pi^2n\left(\lambda-u\right)}
                         \varepsilon \right)
               \sinh(\frac{2\pi^2n}\varepsilon)}
              {n\sinh(\frac{2\pi^2nL}\varepsilon)
               \sinh\left(\frac{2\pi^2n\left(L+2\lambda\right)}
                         \varepsilon \right)},
\end{eqnarray}
\begin{equation}
  f_s(u)=-2\sum_{n=-\infty}^\infty\frac
          {\sinh(\frac{4\pi^2nu}\varepsilon)
          \sinh\left(\frac{4\pi^2n\left(L+2\lambda-2u\right)}
          \varepsilon\right)
          \cosh\left(\frac{2\pi^2n\left(L+3\lambda-1\right)}
          \varepsilon \right)
          \cosh\left(\frac{2\pi^2n\left(\lambda+1\right)}
          \varepsilon \right)}
         {n\sinh(\frac{2\pi^2nL}\varepsilon)
          \cosh\left(\frac{2\pi^2n\left(L+2\lambda\right)}
                    \varepsilon\right)}
\end{equation}
for the bulk and surface free energies. 
Note that in this case we also obtain the free energies of
the level $l$ $A_{2n-1}^{(2)}$ model, which 
corresponds to the level $n$ $C_l^{(1)}$ model
on changing the signs of $u$ and $\lambda$, 
as follows from level-rank duality \cite{k91}.
 
%%%%%%%%%%%%%%%%%%%%%%%%%%%%%%%%%%%%%%%%%%%%%%%%%%%%%%%%%%%%%

We are particularly interested in the critical behavior of these
models as the nome $p=e^{-\varepsilon/L}\to0$. In each case the 
singular term in the free energies is obtained 
by making use of the Poisson summation formula \cite{baxter}.
 
%%%%%%%%%%%%%%%%%%%%%%%%%%%%%%%%%%%%%%%%%%%%%%%%%%%%%%%%%%%%%

{}For the $A_1^{(1)}$ model in regime III we have
\begin{equation}
 f_b \sim \left\{
 \begin{array}{ll}
  p^{2-\alpha_b}\log p & \quad \mbox{for $L=2m$}\\
  \mbox{nsc} & \quad \mbox{for $L=2m+1$}
\end{array}\right.
\label{eqn:a3fb}
\end{equation}
for the bulk, where nsc denotes ``no singular contribution'' and 
$m$ is some integer. The bulk specific heat exponent is given by  
\begin{equation}
 \alpha_b=2-\frac{L}2 .
\label{eqn:a3eb}
\end{equation}
For the surface free energy we find
\begin{equation}
 f_s \sim \left\{
 \begin{array}{ll}
 p^{2-\alpha_s} & \quad \mbox{for $L=2m+1$}\\
p^{2-\alpha_s}\log p & \quad \mbox{for $L=4m$}\\
\mbox{nsc} & \quad \mbox
{for $L=4m+2$}
\end{array}\right.
\label{eqn:a3fs}
\end{equation}
where the excess specific heat exponent is given by 
\begin{equation}
  \alpha_s=2-\frac{L}4 .
\label{eqn:a3es}
\end{equation}

On the other hand, for the $A_1^{(1)}$ model in regime II
we have
\begin{equation}
 f_b \sim \left\{
 \begin{array}{ll}
  p^{2-\alpha_b} & \quad \mbox{for $L\ne 4$}\\
  p^{2-\alpha_b}\log p & \quad \mbox{for $L=4$}
\end{array}\right.
\label{eqn:a2fb}
\end{equation}
for the bulk, with 
\begin{equation}
  \alpha_b=2-\frac{L}{L-2}.
\label{eqn:a2eb}
\end{equation}
For the surface free energy,
\begin{equation}
 f_s \sim \left\{
 \begin{array}{ll}
  p^{2-\alpha_s} & \quad \mbox{for $L\ne 4$}\\
  p^{2-\alpha_s}\log p & \quad \mbox{for $L=4$}
\end{array}\right.
\end{equation}
with 
\begin{equation}
  \alpha_s=2-\frac{L}{2(L-2)}.
\end{equation}  
The bulk results (\ref{eqn:a3fb}), (\ref{eqn:a3eb}), (\ref{eqn:a2fb}),
(\ref{eqn:a2eb}) have been obtained for the ABF model \cite{abf84},
as have the surface results (\ref{eqn:a3fs}) and (\ref{eqn:a3es}) 
\cite{zb96,opb95}.

%%%%%%%%%%%%%%%%%%%%%%%%%%%%%%%%%%%%%%%%%%%%%%%%%%%%%%%%%%%%%

{}For the $B_n^{(1)}$ and $D_n^{(1)}$ models in regime III we have
\begin{equation}
 f_b \sim \left\{
 \begin{array}{ll}
  p^{2-\alpha_b} & \quad
  \mbox{for $L\ne -2m\lambda,L\ne-2m\lambda+1$}\\
  p^{2-\alpha_b}\log p &
  \quad \mbox{for $L=-2m\lambda$}\\
  \mbox{nsc} &
  \quad \mbox{for $L=-2m\lambda+1$}
\end{array}\right.
\end{equation}
with exponent
\begin{equation}
\alpha_b=2+\frac{L}{2\lambda}.
\end{equation}
While for the surface energy, 
\begin{equation}
 f_s \sim \left\{
 \begin{array}{ll}
 p^{2-\alpha_s} & \quad
\mbox{for $L\ne -4m\lambda,L\ne-4m\lambda+\lambda+1$}\\
p^{2-\alpha_s}\log p & \quad \mbox{for $L=-4m\lambda$}\\
\mbox{nsc} & \quad \mbox
{for $L=-4m\lambda+\lambda+1$}
\end{array}\right.
\end{equation}
with exponent
\begin{equation}
\alpha_s=2+\frac{L}{4\lambda}.
\end{equation} 

{}For the $B_n^{(1)}$, $C_n^{(1)}$ and $D_n^{(1)}$ models in 
regime II we have
\begin{equation}
 f_b \sim \left\{
 \begin{array}{ll}
 p^{2-\alpha_b} & \quad
\mbox{for $\frac{L}{L+2\lambda}\ne m_1$ and
      $\frac{2(\lambda+1)}{L+2\lambda}\ne 2m_2-1$}\\[0.4cm]
p^{2-\alpha_b}\log p & \quad
\mbox{for $\frac{L}{L+2\lambda}=m_1$ and
      $\frac{2(\lambda+1)}{L+2\lambda}\ne 2m_2-1$}\\[0.4cm]
\mbox{nsc} & \quad \mbox
{for $\frac{2(\lambda+1)}{L+2\lambda}=2m_2-1$}
\end{array}\right.
\end{equation}
with exponent
\begin{equation}
\alpha_b=2-\frac{L}{L+2\lambda}.
\end{equation} 
For the surface energy
\begin{equation}
 f_s \sim \left\{
 \begin{array}{ll}
 p^{2-\alpha_s} & \quad
\mbox{for $\frac{L}{2(L+2\lambda)}\ne m_1$ and
      $\frac{L+3\lambda-1}{L+2\lambda}\ne 2m_2-1$
      and
      $\frac{\lambda+1}{L+2\lambda}\ne 2m_3-1$}\\[0.4cm]
p^{2-\alpha_s}\log p & \quad
\mbox{for $\frac{L}{2(L+2\lambda)}=m_1$ and
      $\frac{L+3\lambda-1}{L+2\lambda}\ne 2m_2-1$
      and
      $\frac{\lambda+1}{L+2\lambda}\ne 2m_3-1$}\\[0.4cm]
\mbox{nsc} & \quad \mbox
{for  $\frac{L+3\lambda-1}{L+2\lambda}=2m_2-1$ or
      $\frac{\lambda+1}{L+2\lambda}=2m_3-1$}
\end{array}\right.
\end{equation}
with exponent
\begin{equation}
\alpha_s=2-\frac{L}{2(L+2\lambda)}.
\end{equation}
In the above $m_1$, $m_2$ and $m_3$ are arbitrary integers.

In this case the exponents of the level $l$ $A_{2n-1}^{(2)}$ model
follow under level-rank duality with the level $n$ $C_l^{(1)}$ 
model on changing the sign of $\lambda$.

%%%%%%%%%%%%%%%%%%%%%%%%%%%%%%%%%%%%%%%%%%%%%%%%%%%%%%%%%%%%%

The bulk and surface specific heat exponents are seen to
satisfy the relation
$2 \alpha_s = 2 + \alpha_b$.
More generally, this relation can be inferred directly from a comparison
of the singular behaviour of the functional relations for $\kappa_b(u)$ and
$\kappa_s(u)$. 
The known scaling relations \cite{binder} 
$\alpha_b=2-2\nu$, $\alpha_s=\alpha_b+\nu$ are consistent with this
relation and
can be used to infer the value of the correlation length exponent $\nu$.
These relations have been confirmed explicity for the eight-vertex \cite{bz96}
and the CSOS \cite{zb96b} models for which the exponent $\nu$ is known. 
For the present models, in regime III we thus expect   
\begin{equation}
 \nu=-\frac{L}{4\lambda}=\frac{l+g}{2g}=
 \left\{
 \begin{array}{ll}
 \displaystyle{\frac{l+2}{4}}&
\quad \mbox{for $A_1^{(1)}$}\\[0.4cm]
 \displaystyle{\frac{l+2n-1}{2(2n-1)}}&
 \quad \mbox{for $B_n^{(1)}$}\\[0.4cm]
 \displaystyle{\frac{l+2n-2}{4(n-1)}}&
 \quad \mbox{for $D_n^{(1)}$}\\[0.4cm]
\end{array}\right.
\end{equation}
In regime II
\begin{equation}
 \nu=\frac{L}{2(L+2\lambda)}=
      \frac{l+g}{2l}=\left\{
 \begin{array}{ll}
 \displaystyle{\frac{l+2}{2l}}&
\quad \mbox{for $A_1^{(1)}$}\\[0.4cm]
 \displaystyle{\frac{l+2n-1}{2l}}&
 \quad \mbox{for $B_n^{(1)}$}\\[0.4cm]
 \displaystyle{\frac{l+n+1}{2l}}&
 \quad \mbox{for $C_n^{(1)}$}\\[0.4cm]
 \displaystyle{\frac{l+2n-2}{2l}}&
 \quad \mbox{for $D_{n}^{(1)}$}
\end{array}\right.
\end{equation}
and
\begin{equation}
\nu=\frac{L}{2(L-2\lambda)}=\frac{l+n+1}{2n}
\quad \mbox{for $A_{2n-1}^{(2)}$.}
\end{equation}
These results remain to be confirmed via a direct calculation of the
correlation length. 

Our results are consistent with a number of partial checks:

\noindent$\bullet$ Regime III

\noindent (i) There is an equivalence at level $l$ between the $B_1^{(1)}$ 
model and the degree 2 fusion $A_1^{(1)}$ model.
Formally setting $n=1$ in the $B_n^{(1)}$ model 
we see that the bulk free energy of the level $l$ $B_1^{(1)}$ 
model agrees with the result obtained for the level $l$ degree 2 
fusion $A_1^{(1)}$ model \cite{djkmo}. 
The bulk free energy of the $B_n^{(1)}$ model at the critical point ($p \to 0$) is 
consistent with the result obtained from the string hypothesis \cite{kns1,kns2}. 

\noindent (ii) Formally setting $n=2$ in the $D_n^{(1)}$ model, 
the $D_2^{(1)}$ bulk and surface free energies agree with 
twice those of the $A_1^{(1)}$ RSOS model. This is due 
to the fact that $D_2^{(1)}= A_1^{(1)} \oplus A_1^{(1)}$.
The bulk free energy of the $D_n^{(1)}$ model 
at the critical point is also consistent with the result from
the string hypothesis.
Further we can check that the exponent $2/\nu$ of the $D_n^{(1)}$
model is consistent with the result from the thermal scaling relation 
if we identify the dimension of the generalized (1, 3) operator in the conformal field 
theory \cite{ey} as playing the role of the thermal operator. 
  
\noindent$\bullet$ Regime II

\noindent (i) The bulk free energy of the level $l$ $B_1^{(1)}$
model agrees with the degree 2 fusion $A_1^{(1)}$ model with level $l$.

\noindent (ii) The results of the level $l$ $C_1^{(1)}$ model 
are consistent with those of the ($l+1$)-state $A_1^{(1)}$ model. 
The results of the level $1$ $C_n^{(1)}$ model are also 
consistent with those of the 
($n+1$)-state $A_1^{(1)}$ model in regime III with nome $p^2$.
This latter correlation length exponent has been recently obtained  directly for the
ABF model in regime III \cite{op96}.

\noindent (iii) The bulk and surface free energies of the $D_2^{(1)}$ model 
agree with twice those of the $A_1^{(1)}$ RSOS model.\\

\noindent
{\bf Acknowledgements}\\
MTB and YKZ are supported by the Australian Research Council.
VF is supported by an Australian Postgraduate Research Award.
%%%%%%%%%%%%%%%%%%%%%%%%%%%%%%%%%%%%%%%%%%%%%%%%%%%%%%%%%

%%%%%%%%%%%%%%%%%%%%%%%%%%%%%%%%%%%%%%%%%%%%%%%%%%%%%%%%%
\vfill
\begin{tabular}{@{}*{8}{c}}
{\bf Table 1}\\  \hline \\

type &  $A_1^{(1)}$ & $B_n^{(1)}$ $(n\ge2)$ & $C_n^{(1)}$ $(n\ge1)$  &
$D_n^{(1)}$ $(n\ge3)$ \\[2.0ex] 

level & $l \ge 2$ & $l \ge 2$ & $l \ge 1$ & $l \ge 2$ \\[2.0ex] 

$g$   & $n+1$ & $2n-1$  & $n+1$
 & $2n-2$  \\[2.0ex]

$t$ & 1 & 1 & 2 & 1 \\[2.0ex]

$\lambda$ & $-1$ & $-n+\frac{1}{2}$ & $-n-1$ & $-n+1$ \\[2.0ex]

$L$ & $l+2$  & $l+2n-1$ & $2(l+n+1)$ & $l+2n-2$ \\[2.0ex] \hline

\end{tabular}
%%%%%%%%%%%%%%%%%%%%%%%%%%%%%%%%%%%%%%%%%%%%%%%%%%%%%%%%%
\end{document}